\documentclass[twocolumn,aps,amsmath,showpacs,superscriptaddress]{revtex4-1}

\usepackage{graphicx,dcolumn,bm,amssymb,amsmath,ulem,indentfirst,amsthm}

\usepackage{graphicx,dcolumn,bm,amssymb,amsmath,ulem,indentfirst,amsthm,caption,subcaption}

\usepackage[toc,page]{appendix}

\usepackage{color}

\theoremstyle{plain}

\def\be{\begin{equation}}

\def\ee{\end{equation}}

\newtheorem*{theorem*}{Theorem}

\begin{document}

\author{Bingyu Cui}

\affiliation{Cavendish Laboratory, University of Cambridge, JJ Thomson Avenue, CB3 0HE Cambridge, U.K.}

\author{Alessio Zaccone}

\email{az302@cam.ac.uk}

\affiliation{Department of Physics ``A. Pontremoli", University of Milan, via Celoria 16, 20133 Milano, Italy}

\affiliation{Cavendish Laboratory, University of Cambridge, JJ Thomson Avenue, CB3 0HE Cambridge, U.K.}

\affiliation{Statistical Physics Group, Department of Chemical Engineering and Biotechnology, University of Cambridge, Philippa Fawcett Drive, CB3 0AS Cambridge, U.K.}

\author{David Rodney}

\email{david.rodney@univ-lyon1.fr}

\affiliation{Institut Lumi\`ere Mati\`ere, UMR 5306 Universit\'e Lyon 1-CNRS, Universit\'e de Lyon, F-69622 Villeurbanne, France}

\begin{abstract}

A lattice dynamical formalism based on nonaffine response theory is derived for non-centrosymmetric crystals, accounting for long-range interatomic interactions using the Ewald method. The framework takes equilibrated static configurations as input to compute the elastic constants in excellent agreement with both experimental data and calculations under strain. Besides this methodological improvement, which enables faster evaluation of elastic constants without the need of explicitly simulating the deformation process, the framework provides insights into the nonaffine contribution to the elastic constants of $\alpha$-quartz. It turns out that, due to the non-centrosymmetric lattice structure, the nonaffine (softening) correction to the elastic constants is very large, such that the overall elastic constants are at least 3-4 times smaller than the affine Born-Huang estimate.

\end{abstract}


\title{Nonaffine lattice dynamics with the Ewald method reveals strongly nonaffine elasticity of $\alpha$-quartz}

\maketitle

\section{Introduction}

\label{Intro}

Lattice dynamics has been formulated through the pioneering work of Max Born and co-workers on the simplifying assumption that deformations are homogeneous, or in modern language, affine~\cite{Born}. In practice, this implies that every atom is displaced under deformation by the macroscopic strain tensor operating on the original position vectors. This transformation defines the affine positions in the deformed lattice. Such a description assumes however that mechanical equilibrium is satisfied at the affine positions, which is certainly true for centrosymmetric lattices, where, owing to each atom being a local center of inversion symmetry, the forces transmitted by the neighbours cancel out by symmetry at the affine  positions.

The situation is different for disordered lattices like glasses and for non-centrosymmetric crystals as well as near crystalline defects like grain boundaries. In such cases, the atoms are not centers of symmetry in their affine positions and therefore receive from their neighbours forces that sum up to a net force. The latter is released via an extra displacement, called nonaffine displacement or relaxation, which brings the atoms to final positions that do not coincide with the affine positions.

Reformulating the equations of motion by explicitly requiring that the atoms move along nonaffine pathways of mechanical equilibrium (where the net force on each atom is zero at all steps) leads to a negative (softening) correction to the elastic constants, which was first expressed analytically by Lemaitre and Maloney~\cite{Lemaitre} for systems of particles, which interact through short-ranged pairwise potentials.

The resulting framework is known as the nonaffine response theory or nonaffine lattice dynamics and has recently been applied to various systems and materials, from packings (where it recovers the $\sim(z-2d)$ jamming scaling, with $z$ the coordination number~\cite{Scossa-Romano}) to polymers~\cite{Zaccone2013}, and to analyze dissipation in high-frequency oscillatory rheology~\cite{Damart}. The framework also provides quantitative predictions of dynamic viscoelastic moduli of coarse-grained (Kremer-Grest) glassy polymers~\cite{Palyulin}.

While the effects of nonaffinity have been intensively investigated in glassy materials, the same is not true for crystals. Here we show that nonaffine effects are very strong in a prototypical non-centrosymmetric crystal: $\alpha$-quartz, for which the non-centrosymmetry is also the root cause of piezoelectricity \cite{Kholkin}. Interatomic interactions are modeled using a classical BKS potential \cite{BKS}, which includes both a short-ranged potential and long-ranged Coulomb interactions between partial charges on silicon and oxygen atoms. In disordered glasses, these interactions can be treated with a truncation \cite{Wolf,Fennell06,Carre07} and can therefore be handled using the original approach of Lemaitre and Maloney \cite{Damart}. By way of constrast, in an ordered crystal like $\alpha$-quartz, Coulomb interactions are conditionally convergent and must be treated accordingly using the Ewald method~\cite{Ewald,Lee}. We show below that the corresponding long-ranged many-body contribution can be also treated analytically and incorporated in the nonaffine response theory to predict the elastic constants of $\alpha$-quartz.

Surprisingly, we find that nonaffine contributions are not small corrections to the elastic constants: they are substantial (negative) contributions, which make the resulting elastic constants up to 4 times smaller than the affine estimates. This important fact has been overlooked in previous studies on $\alpha$-quartz lattice dynamics~\cite{Bosak}.

These results are also relevant to studies of the boson peak (i.e. excess with respect to Debye's $\omega^{2}$ law in the vibrational density of states), which is typically observed in glasses but has recently also been measured in $\alpha$-quartz~\cite{Chumakov}. Recent works have highlighted the close connection between nonaffine elasticity and the boson peak anomaly~\cite{Milkus,Sengupta}, and it has been suggested that the root cause of both boson peak and nonaffine elasticity could be traced back to the inherent lack of centrosymmetry of both glasses and non-centrosymmetric crystals such as $\alpha$-quartz~\cite{Milkus}.



\section{Atomistic simulations of $\alpha$-quartz crystal}
\subsection{Lattice structure}
\label{LattStruc}
\begin{figure*}	
\centering
	\begin{subfigure}[t]{1.5in}
		\caption{}
		\includegraphics[width=1.5in]{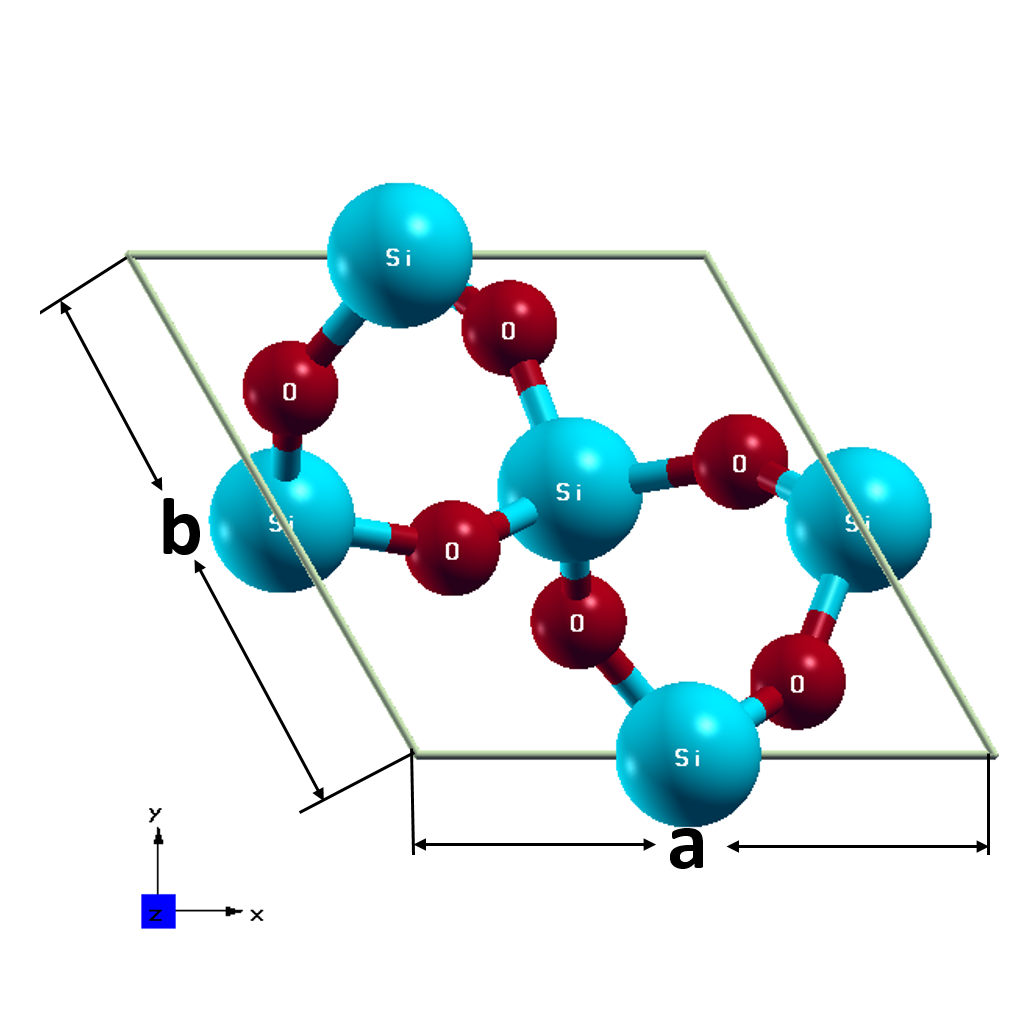}				
	\end{subfigure}
	\begin{subfigure}[t]{1.5in}
		\caption{}
		\includegraphics[height=1.5in]{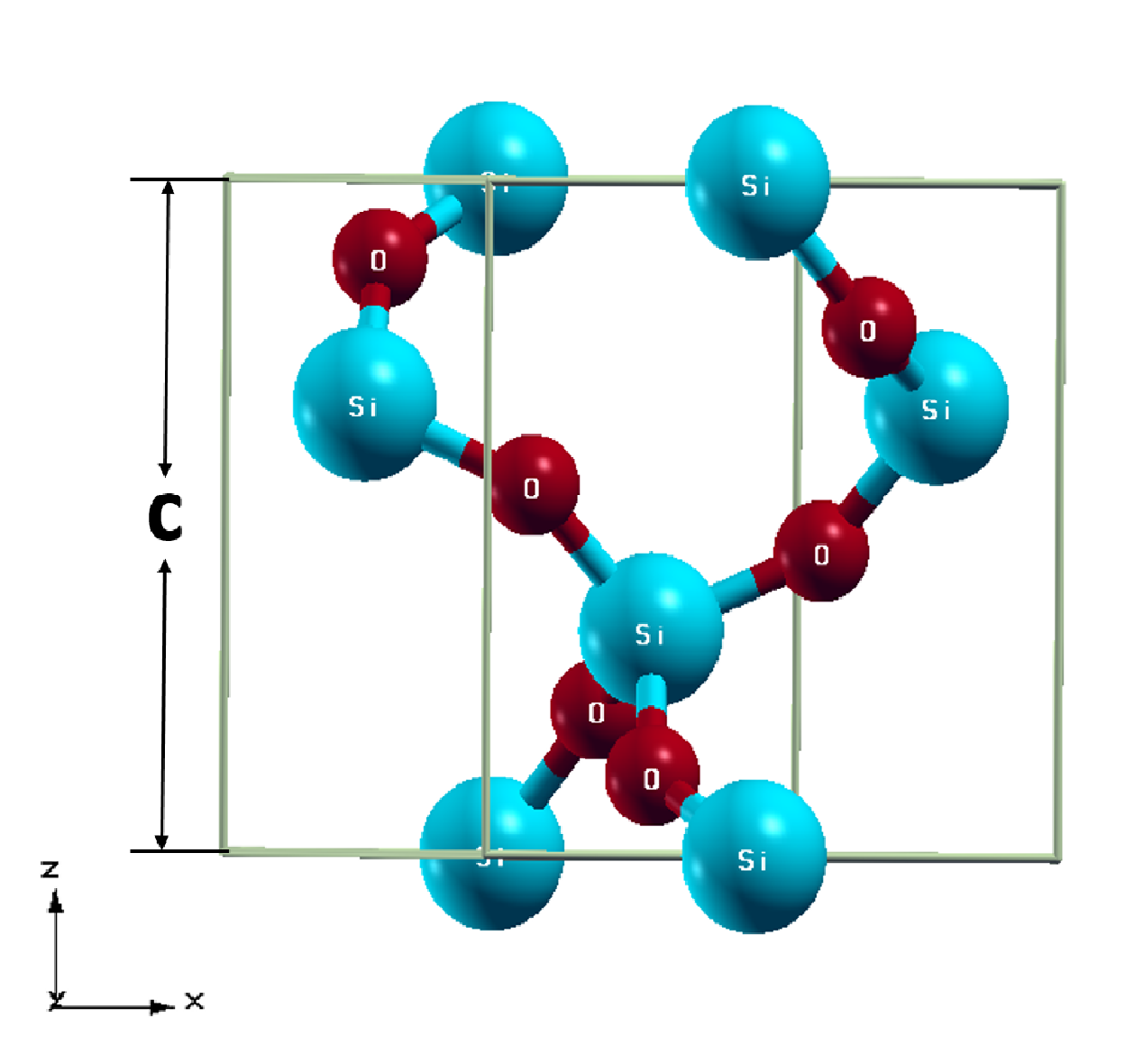}		
	\end{subfigure}
	\begin{subfigure}[t]{1.5in}
		\caption{}
		\includegraphics[height=1.5in]{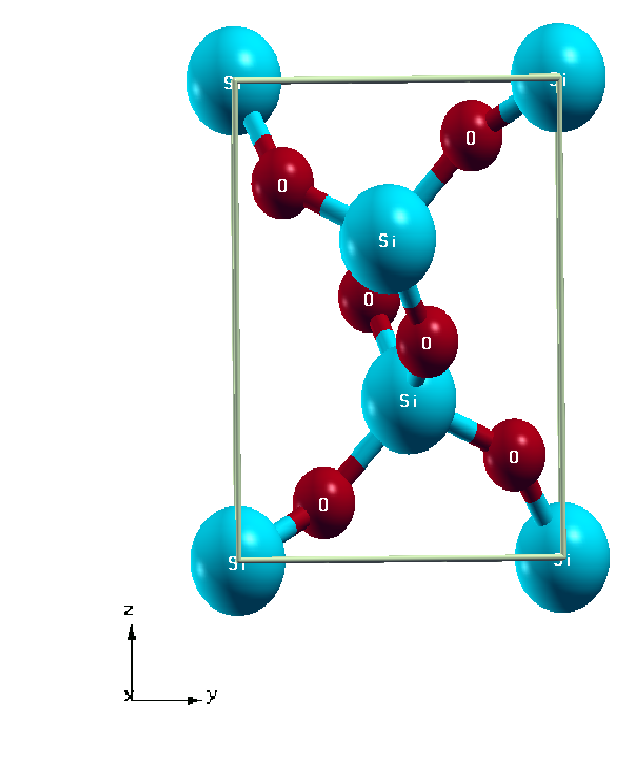}		
	\end{subfigure}






\caption{Unit cell of $\alpha$-quartz from different perspectives: (a) top view (b) left view (c) front view. Si atoms are in cyan, O atoms in red.
}
\label{quartzview}
\end{figure*}

X-ray and neutron crystallography have been applied to many materials to determine the crystal
structure and atomic positions, including $\alpha$-quartz. It has been shown that crystals of $\alpha$-quartz have a trigonal Bravais lattice  composed of $\text{SiO}_4$ tetrahedra that are linked together at their corners to form a three-dimensional network \cite{Sutter}. The conventional unit cell, shown in Fig. \ref{quartzview}, is hexagonal and contains three molecules of $\text{SiO}_2$. Its c-axis is a threefold screw axis; that is,  the lattice remains unchanged after a rotation of $120^\circ$ about this axis followed by a translation of $+c/3$ along the same axis. Along the negative $c$ direction, the screw axis is left-handed if the $120^\circ$ rotation appears clockwise while if the rotation appears counterclockwise, the screw axis is right-handed. $\alpha$-quartz may exist in either of these forms, which are enantiomorphs (mirror images). $\alpha$-quartz crystals rotate the polarization of light propagating parallel to the $c$-axis, which is therefore also called the optical axis, in the same sense as the screw. Perpendicular to the $c$-axis, are three twofold axes that are separated from one another by angles of $120^\circ$ and intercept the c-axis at intervals of $c/3$. The absence of an inversion center allows $\alpha$-quartz to exhibit piezoelectricity when pressed along one of the twofold axes that are therefore often named electrical axes \cite{Sutter}.

Two space groups, $P3_121$ or $P3_221$, can be used to label the $\alpha$-quartz, depending on whether the $c$-axis is left- or right-handed. In this paper, we initially used the consistent results of lattice constants from Bragg \& Gibbs \cite{Bragg}, Wyckoff \cite{Wyckoff} and Kihara \cite{Kihara}, with lattices parameters $a$ and $c$ at $298$K equal to $4.9137\text{\AA}$ and $5.4047\text{\AA}$ respectively. The atomic positions of left-handed $\alpha$-quartz are given in the right-handed hexagonal coordinate systems in Table \ref{UnitCell}.
\begin{table}[tbp]

\centering

\begin{tabular}{lccc}

\hline

Atom &x &y &z\\ \hline

Si &0.4697 &0 &0\\

Si &0 &0.4697 &2/3\\

Si &0.5303 &0.5303 &1/3\\

O &0.4133 &0.2672 &0.1188\\

O &0.2672 &0.4133 &0.5479\\

O &0.7328 &0.1461 &0.7855\\

O &0.5867 &0.8539 &0.2145\\

O &0.8539 &0.5867 &0.4521\\

O &0.1461 &0.7328 &0.8812

\end{tabular}

\caption{Fractional coordinates of atoms of left-handed $\alpha$-quartz given in the scaled unit at 298K at ambient pressure \cite{Kihara}.}
\label{UnitCell}
\end{table}

\subsection{Empirical potential}
In the present work, the cohesion of $\alpha$-quartz is modeled with the classical BKS potential, which is based on a short-range Buckingham potential and long-range Coulombic interactions between partial charges on Si and O atoms. Different parametrizations of this potential exist \cite{BKS,Mantisi}. We have used the original parameters \cite{BKS}, which do not include any direct Si-Si interaction, because they provide the best agreement with experimental measurements of elastic constants of $\alpha$-quartz \cite{Carre}. The short-ranged potential between atoms $i$ and $j$ is expressed as:

\begin{align}
\Phi_{ij}^{sh}(r)=&\left\{A_{ij}e^{-\frac{r}{\rho_{ij}}}-\frac{C_{ij}}{r^6}-\left[A_{ij}e^{-\frac{r_{c,sh}}{\rho_{ij}}}-\frac{C_{ij}}{r^6_{c,sh}}\right]\right\}\nonumber\\
&\times\Theta(r_{c,sh}-r),
\label{BKS}
\end{align}
where $\Theta(r)$ is the Heaviside function and $r$ the distance between atoms. The parameters of the potential are given in Table \ref{paramBKS}. The best agreement with experimental data is obtained for $r_{c,sh}=10\text{\AA}$~\cite{Carre}.

In order to treat the Coulombic interactions analytically, we used the classical Ewald method \cite{Ewald,Born,Toukmaji96,Lee}. In this approach, the point charge distribution, which is described by delta functions, is transformed by adding and subtracting Gaussian distributions. The total electrostatic energy is then re-written as the sum of a short-range term (difference between point- and gaussian charge distributions) in real space, a long-range term (Gaussian charge distribution) in Fourier space plus a self-interaction constant:

\begin{align}
E&\equiv E_{SR}+E_{LR}+E_{SI}\nonumber\\
&=\frac{1}{4\pi\epsilon_0}\frac{1}{2}\sum
_{i\neq j}\frac{q_iq_j}{r_{ij}}\text{erfc}(\frac{r_{ij}}{\sqrt{2}\sigma})\nonumber\\
&+\frac{1}{2V\epsilon_0}\sum_{\mathbf{G}\neq\mathbf{0}}\frac{\exp(-\sigma^2G^2/2)}{G^2}|S(\mathbf{G})|^2\nonumber\\
&-\frac{1}{4\pi\epsilon_0}\frac{1}{\sqrt{2\pi}\sigma}\sum_iq_i^2,
\label{Etot}
\end{align}

where $q_i$ is the charge on atom $i$, $\text{erfc(z)}=1-2/\sqrt{\pi}\int_0^z\exp(t^2)dt$ is the complementary error function, $\mathbf{G}=2\pi[n_x/L_x,n_y/L_y,n_z/L_z]$ refers to reciprocal lattice vectors and $S(\mathbf{G})=\sum_jq_j\exp(i\mathbf{G}\cdot\mathbf{R}_j)$ is the structure factor. Here, $L_x,L_y,L_z$ are the dimensions of the simulation cell, which is assumed periodic and orthogonal. The parameter $\sigma$ is the standard deviation of the Gaussian distribution. It sets the cross-over between the real and reciprocal terms, which both converge absolutely and rapidly. In the literature, one may also write $\alpha=1/\sqrt{2}\sigma$. It is recommended for accurate calculations to use a cut-off radius for the real space potential $R_{cut}=3.12/\alpha$ and a summation in reciprocal space up to $n_{\kappa,max}=\alpha L_\kappa$. We used $R_{cut} = 10\text{\AA}$, which is a  trade-off between the computing times of the short-ranged term, $E_{SR}$, and of the long-range summation in Fourier space, $E_{LR}$.

In the following, the short-range and self-interaction terms will be included in the short-range BKS term of Eq. \ref{BKS}. This term can be treated with the original approach of Lemaitre and Maloney \cite{Lemaitre}, which is summarized in Section \ref{pairpot}. Only $E_{LR}$ requires a special treatment because of its many-body nature, as detailed in Section \ref{Sec:Ewald}.

\begin{table}
\centering
\begin{tabular}{lccc}
\hline
 &$A_{ij}$(eV) &$\rho_{ij}(\AA)$ &$C_{ij}(\text{eV} \AA)$ \\ \hline
O-O &1388.773 &0.3623 &175.0\\
Si-O &18003.7572 &0.2052 &133.5381\\
\end{tabular}
\caption{Parameters of the empirical potential used to model $\alpha$-quartz.}
\label{paramBKS}
\end{table}

\subsection{Simulation procedure}

Since we consider the properties of a perfect crystal, the system can in principle be limited to a single unit cell. In practice, we used a small but finite system, containing 1350 atoms in a periodic orthogonal cell. We started from the lattice positions in Table \ref{UnitCell} and the experimental lattice constants \cite{Bragg,Wyckoff,Kihara}. We then relaxed the simulation cell at 0K by energy minimization, adapting the cell dimensions with a barostat to impose zero internal stresses. The equilibrium lattice constants thus obtained are $a=4.94$ and $c=5.44$ \AA, corresponding to a density of 2.60 g/cm$^3$, close to the experimental value of 2.65 g/cm$^3$ \cite{Heyliger03,Wang2015}.

To validate the analytical expressions of the elastic constants, we computed numerically their values by straining the crystal in small increments (1e-5) and computing the slope of the resulting stress-strain curves. To obtain the affine constants, no relaxation was allowed between affine deformation steps, i.e. the atoms remained at their affine positions, while the nonaffine constants were computed by relaxing the atomic positions at fixed cell shape between each strain increment.

\section{Nonaffine lattice dynamics formalism with the Ewald method}

\subsection{Contribution of pairwise potential}

\label{pairpot}

We start by summarizing the expressions of the affine and nonaffine elastic constants in the case of particules that interact through a short-ranged pairwise potential, $V_{ij}(r)$. We consider a system of $N$ atoms of mass $\{m_i\}$ in a volume $V$. Defining $t_{ij}=\frac{\partial V_{ij}}{\partial R_{ij}}$ and $c_{ij}=\frac{\partial^2 V_{ij}}{\partial R_{ij}^2}$, one can show \cite{Lemaitre,Scossa-Romano,Damart} that the elastic constants are written as the difference between the affine (or Born) elastic constants and nonaffine terms:

\begin{equation}
C_{\alpha\beta\kappa\chi} = C^{Born}_{\alpha\beta\kappa\chi}-\frac{1}{V}\sum_{m=1}^{3N-3}\frac{C_{m,\alpha\beta}C_{m,\kappa\chi}}{\omega_m^2}.
\label{ElastConst}
\end{equation}

The affine elastic constants are expressed as:

\begin{eqnarray}
	C^{Born}_{\alpha\beta\kappa\chi} &=& -\frac{1}{4V} \sum_{i \neq j} [D_{ij}^{\alpha\kappa} R_{ij}^\beta + D_{ij}^{\beta\kappa} R_{ij}^\alpha ] R_{ij}^\chi \\ \nonumber
	&=&\frac{1}{2V}\sum_{i \neq j}(R_{ij}c_{ij}-t_{ij})R_{ij}n_{ij}^\alpha n_{ij}^\beta n_{ij}^\kappa n_{ij}^\chi,
\end{eqnarray}

where $D_{ij}^{\alpha \beta}$ is the dynamical matrix of the system, $\textbf{R}_{ij}$ the vector between atoms $i$ and $j$ and $\textbf{n}_{ij}$ the corresponding unit vector. The nonaffine term is written as a sum over the normal modes $m$ of the system with $C_m$, a mode-dependent tensor expressed as:

\begin{equation}
	C_{m,\kappa\chi} = -\sum_{j\alpha} \Xi^\alpha_{j,\kappa\chi} \frac{e^\alpha_j(m)}{\sqrt{m_j}},
	\label{Cm}
\end{equation}

where $e^\alpha_j(m)$ is the component on atom $j$ and direction $\alpha$ of the m$^{th}$ eigenvector of the mass-scaled dynamical matrix of the system. The corresponding eigenfrequency, $\omega_m$, appears in Eq. \ref{ElastConst}. $\Xi^\alpha_{j,\kappa\chi}$ is the nonaffine force vector field, which corresponds to the force that appears on the atoms when an incremental affine deformation $d \epsilon_{\kappa\chi}$ is applied to the system. This force drives nonaffine relaxations. For a pair potential, we have:

\begin{align}
	\Xi_{i,\kappa\chi}^\alpha &= \frac{\partial F_i^\alpha}{\partial \epsilon_{\kappa\chi}} = -\frac{1}{2} \sum_j (D_{ij}^{\alpha\kappa}R_{ij}^\beta + D_{ij}^{\alpha\beta}R_{ij}^\kappa ) \notag\\
	&=-\sum_j(R_{ij}c_{ij}-t_{ij})n_{ij}^\alpha n_{ij}^\kappa n_{ij}^\chi.
	\label{Xi_shortrange}
\end{align}








\subsection{Ewald sum contribution}

\label{Sec:Ewald}

We now consider the contribution of the long-ranged term $E_{LR}$ in Eq. \ref{Etot} to the affine and nonaffine elastic constants. The expressions in Eqs. \ref{ElastConst} and \ref{Cm} remain valid but we need to express the contribution of $E_{LR}$ to the dynamical matrix, the affine elastic constants and the nonaffine forces.

\subsubsection{Forces and dynamical matrix}

The long-ranged energy $E_{LR}$ produces atomic forces due to the dependence of the structure factor, $S(\mathbf{G})=\sum_jq_j\exp(i\mathbf{G}\cdot\mathbf{R}_j)$, on atomic positions. The expression of the resulting force is \cite{Toukmaji96,Lee}:

\begin{align}
\mathbf{F}_i &= -\frac{\partial E_{LR}}{\partial\mathbf{R}_i}\nonumber\\
&=-\frac{1}{2V\varepsilon_0}\sum_{\mathbf{G}\neq\mathbf{0}}\frac{\exp{(-\sigma^2G^2/2)}}{G^2}[S(\mathbf{G})(-i\mathbf{G})q_ie^{-i\mathbf{G\cdot R}_i}\nonumber\\
&+S(-\mathbf{G})q_i(i\mathbf{G})e^{i\mathbf{G\cdot R}_i})]\nonumber\\
&=-\frac{1}{V\varepsilon_0}\sum_{\mathbf{G}\neq\mathbf{0}}\frac{\exp{(-\sigma^2G^2/2)}}{G^2}\mathbf{G} q_i Im[S(\mathbf{G})e^{-i\mathbf{G\cdot R}_i}]\nonumber\\
&=\frac{q_i}{V\varepsilon_0}\sum_{\mathbf{G}\neq\mathbf{0}}\frac{\exp{(-\sigma^2G^2/2)}}{G^2}\mathbf{G}\sum_jq_j\sin{(\mathbf{G}\cdot\mathbf{R}_{ij})}.
\end{align}

In the following, we simplify the notations by noting $I(u) = \exp(-\sigma^2u/2)/u$, such that the contribution of the Ewald long-range term to the atomic force is written as:

\begin{equation}
	\mathbf{F}_i = \frac{q_i}{V\varepsilon_0}\sum_{\mathbf{G}\neq\mathbf{0}}I(G^2)\mathbf{G}\sum_jq_j\sin{(\mathbf{G}\cdot\mathbf{R}_{ij})}
	\label{EwaldForce}
\end{equation}

The long-range contribution to dynamical matrix elements can be computed likewise:\\

1. $i\neq j$:

\begin{align*}
D_{ij}^{\alpha\beta}&=\frac{\partial^2E_{LR}}{\partial R_{i}^{\alpha}\partial R_j^{\beta}}\\ \nonumber
&=\frac{q_iq_j}{V\varepsilon_0}\sum_{\mathbf{G}\neq\mathbf{0}}I(G^2)G^{\alpha}G^{\beta}\cos{(\mathbf{G}\cdot\mathbf{R}_{ij})}
\end{align*}
\begin{equation}
\label{dyna}
\end{equation}

2. $i=j$:

\begin{align*}
D_{ii}^{\alpha\beta}&=-\frac{q_i}{V\varepsilon_0}\sum_{\mathbf{G}\neq\mathbf{0}}I(G^2)G^{\alpha}G^{\beta}\sum_{j\neq i}q_j\cos(\mathbf{G}\cdot\mathbf{R}_{ij})\\ \nonumber
&=-\sum_{j\neq i}D_{ij}^{\alpha\beta}
\end{align*}

\subsubsection{Tensile deformation}

To find the long-range effect on the nonaffine forces, we need to express the variation of the atomic force in Eq. \ref{EwaldForce} when an incremental affine strain is applied to the system. We consider first a uniaxial strain $\epsilon$ along direction $x$. The dependence on $\epsilon$ is due to the dependence of three terms:

\begin{itemize}
	\item the volume, $V \rightarrow V(1+\epsilon)$
	\item the reciprocal vectors, which in an orthogonal box become $\mathbf{G} \rightarrow 2\pi[n_X/L_X(1+\epsilon),n_Y/L_Y,n_Z/L_Z]$
	\item the atom-to-atom vectors, $\mathbf{R}_{ij} \rightarrow [R_{ij}^x(1+\epsilon),R_{ij}^y,R_{ij}^z]$
\end{itemize}

We note that with these transformations, $\mathbf{G}.\mathbf{R}_{ij}$ is unchanged and so that the structure factor $S(\mathbf{G})$ is constant. Taking the derivative of $\mathbf{F}_i$ in Eq. \ref{EwaldForce} with respect to $\epsilon$ and we obtain in the limit $\epsilon \rightarrow 0$:

\begin{equation}
\mathbf{\Xi}_{i,xx}=-\frac{q_i}{V\varepsilon_0}\sum_{\mathbf{G}\neq\mathbf{0}}I(G^2)(\sigma^2+\frac{2}{G^2})G_x^2\mathbf{G}\sum_j q_j \sin{(\mathbf{G}\cdot\mathbf{R}_{ij})}.
\label{tensileXi}
\end{equation}

Taking the first and second derivatives of $E_{LR}$ with respect to $\epsilon$, we obtain the tensile stress and affine elastic constants for the tensile strain:

\begin{align}
	\sigma_{xx} &=\lim_{\epsilon \rightarrow 0} \frac{1}{V} \frac{\partial E_{LR}}{ \partial \epsilon} \notag\\
	&= \frac{1}{2 V^2 \epsilon_0} \sum_{\mathbf{G}\neq\mathbf{0}}I(G^2) |S(\mathbf{G})|^2 \Big( [\sigma^2 + \frac{2}{G^2}]G_x^2 - 1 \Big)
\end{align}

and

\begin{align}
	C^{Born}_{xxxx}&=\lim_{\epsilon \rightarrow 0} \frac{1}{V} \frac{\partial^2 E_{LR}}{\partial \epsilon^2} =\frac{1}{V^2\epsilon_0}\sum_{\mathbf{G}\neq\mathbf{0}}I(G^2)|S({\mathbf{G}})|^2\times\nonumber\\
&\left(1-\frac{5}{2}[\sigma^2+\frac{2}{G^2}]G_{x}^2 + [\frac{4}{G^4}+2\frac{\sigma^2}{G^2} + \frac{\sigma^4}{2}]G_{x}^4\right).
\end{align}
Similar expressions are obtained for tensile deformations along $y$ and $z$. Finally, the cross-terms are expressed as:

\begin{align}
	C^{Born}_{\alpha \alpha \kappa\kappa}&=\frac{1}{V^2\epsilon_0}\sum_{\mathbf{G}\neq\mathbf{0}}I(G^2)|S({\mathbf{G}})|^2\times\nonumber\\
&\left(\frac{1}{2} - [\sigma^2+\frac{2}{G^2}]\frac{G_\alpha^2 + G_\kappa^2}{2} + [\frac{4}{G^4}+2\frac{\sigma^2}{G^2} + \frac{\sigma^4}{2}] G_\alpha^2 G_\kappa^2 \right)
\end{align}
for Cartesian components $\alpha,\kappa$.

\subsubsection{Shear deformation}

We now consider the case of an affine shear strain parallel to the $y$ planes with displacements along the $x$ direction. The applied strain is noted $\gamma_{xy} \equiv \gamma$. Under this strain, the axis of the box become: $\mathbf{a}_1^\prime=(L_x,0,0)=\mathbf{a}_1, \mathbf{a}_2^\prime=(L_x\gamma,L_y,0), \mathbf{a}_3^\prime=(0,0,L_z)=\mathbf{a}_3$ while the reciprocal vectors become: $\mathbf{G}^\prime=2\pi(\frac{n_x}{L_x},\frac{n_y}{L_y}-\frac{n_x\gamma}{L_x},\frac{n_z}{L_z})$ and the atom-to-atom vectors become: $\mathbf{R}_{ij}^\prime=(R_{ij}^x+R_{ij}^y\gamma,R_{ij}^y,R_{ij}^z)$. One can check that again $\mathbf{G}\cdot\mathbf{R}_{ij}$ is unchanged during the transformation. After taking the derivative of the long-range force in Eq. \ref{EwaldForce} with respect to $\gamma$, we obtain in the limit $\gamma \rightarrow 0$:

\begin{align}
\mathbf{\Xi}_{i,xy}&=-\frac{q_i}{V\varepsilon_0}\sum_{\mathbf{G}\neq\mathbf{0}}I(G^2)(\sigma^2 + \frac{2}{G^2}) \times \\ \nonumber
&G_xG_y \mathbf{G} \sum_j q_j \sin{(\mathbf{G}\cdot\mathbf{R}_{ij})}.
\label{Xi_longrange_shear}
\end{align}

Similarly, the shear stress is expressed as:

\begin{align}
\sigma_{xy}=\frac{1}{2V^2\varepsilon_0}\sum_{\mathbf{G^\prime}\neq\mathbf{0}}I(G^2)|S(\mathbf{G})|^2 ( \sigma^2 + \frac{2}{G^2} ) G_xG_y
\end{align}

and the affine elastic constant:

\begin{align}
C^{Born}_{xyxy}&=\lim_{\gamma\rightarrow0}\frac{\partial\sigma_{xy}}{\partial\gamma}=\frac{1}{2V^2\varepsilon_0}\sum_{\mathbf{G}\neq\mathbf{0}}I(G^2)S({\mathbf{G}})|^2\times\nonumber\\
&\left(\sigma^4+4\frac{\sigma^2}{G^2}+\frac{8}{G^4}\right)G_x^2G_y^2
\end{align}

\subsubsection{Some other affine elastic constants from $E_{LR}$}
We note that the formula of Born approximation hold for a generic strain tensor $\underline{\underline{\eta}}$ \cite{Lemaitre}:
\begin{equation}
C^{Born}_{\alpha\beta\kappa\chi}= \lim_{\underline{\underline{\eta}} \rightarrow 0} \frac{1}{V}\frac{\partial E_{LR}}{\partial\eta_{\alpha\beta}\partial\eta_{\kappa\chi}}
\end{equation}
For $C_{16}=C_{xxxy}$, $C_{14}=C_{xxyz}$ and $C_{56}=C_{xyxz}$, we have respectively,
\begin{widetext}
\begin{align}
C_{xxxy}&=\frac{1}{2V\epsilon_0}\sum_{\mathbf{G}\neq\mathbf{0}}I(G^2)|S(\mathbf{G})|^2\left[(\sigma^4+\frac{4\sigma^2}{G^2}
+\frac{4}{G^4})G_x^2-(\sigma^2+\frac{2}{G^2})\right]G_xG_y+\frac{1}{2V\epsilon_0}\sum_{\mathbf{G}\neq\mathbf{0}}I(G^2)|S(G^2)|^2\frac{2G_x^2}{G^4}2G_xG_y\notag\\
&=\frac{1}{2V\epsilon_0}\sum_{\mathbf{G}\neq\mathbf{0}}I(G^2)|S(G^2)|^2\left[(\sigma^4+\frac{4\sigma^2}{G^2}+\frac{8}{G^4})G_x^2-\sigma^2-\frac{2}{G^2}\right]G_xG_y\\
C_{xxyz}&=\frac{1}{2V\epsilon_0}\sum_{\mathbf{G}\neq\mathbf{0}}I(G^2)|S(G^2)|^2\left[(\sigma^4+\frac{4\sigma^2}{G^2}+\frac{8}{G^4})G_x^2-\sigma^2-\frac{2}{G^2}\right]G_yG_z\\
C_{xyxz}&=\frac{1}{2V^2\epsilon_0}\sum_{\mathbf{G}\neq\mathbf{0}}I(G^2)|S(G^2)|^2\left(\sigma^4+\frac{4}{G^4}+\frac{4\sigma^2}{G^2}\right)G_xG_yG_xG_z
+\frac{1}{2V^2\epsilon_0}\sum_{\mathbf{G}\neq\mathbf{0}}I(G^2)|S(G^2)|^2\frac{4}{G^4}G_xG_yG_xG_z\notag\\
&=\frac{1}{2V^2\epsilon_0}\sum_{\mathbf{G}\neq\mathbf{0}}I(G^2)|S(G^2)|^2\left(\sigma^4+\frac{8}{G^4}+\frac{4\sigma^2}{G^2}\right)G_xG_yG_xG_z
\end{align}
\end{widetext}

\section{Results and Discussion}

\begin{figure}[h]	
\centering

 \includegraphics[width=2.5in]{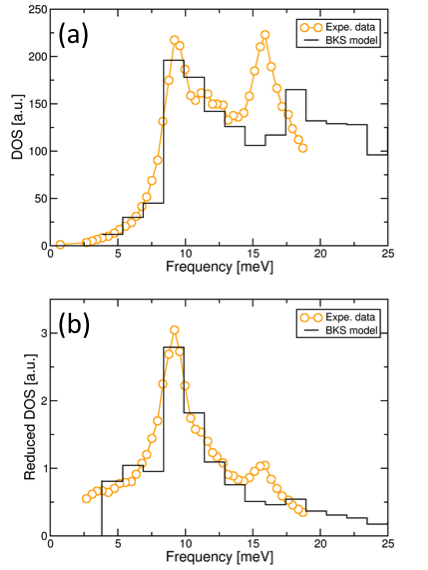}
	\caption{Comparison of the density of states (a) and reduced density of states (DOS normalized by the frequency squared) (b) obtained numerically with the BKS model and experimentally with inelastic x-ray scattering \cite{Chumakov}.}
\label{DOS}
\end{figure}

We first use Eqs. \ref{dyna} to compute the dynamical matrix of the present atomic-scale model of $\alpha$-quartz and, after diagonalization, obtain the vibrational density of states (VDOS). The result is shown in Fig. \ref{DOS}(a), with a comparison to the experimental data obtained by Chumakov \textit{et al} \cite{Chumakov}. The present implementation of the BKS model predicts accurately the first peak of the VDOS, which occurs at about 10 meV. The second peak is reproduced only qualitatively, being located at a slightly higher frequency (18 instead of 16 meV) and with a slightly lower amplitude. Normalizing the VDOS by $\omega^2$ in Fig. \ref{DOS}(b), we see that the numerical model reproduces well the boson peak reported experimentally. We can conclude that the present model reproduces satisfactorily the VDOS and boson peak of $\alpha$-quartz.

Second, we use Eq. \ref{ElastConst} with the short- and long-ranged terms presented above to compute both the affine and total elastic constants of $\alpha$-quartz. The result is given in Table III with a comparison to experimental data. We checked by direct numerical calculations that the analytical expressions described in previous Section predict faithfully the elastic constants. We chose the same parametrization of the BKS potential and Ewald summation as Carr\'e \textit{et al} \cite{Carre}, because they yield a very good agreement with experimental data, as seen in Table III, when both the affine and nonaffine contributions are included. On the other hand, when only the affine deformation is allowed, the elastic constants are largely overestimated, by a factor 3 to 4 for $C_{11}$, $C_{33}$, $C_{44}$, $C_{66}$, and up to a factor of 15 for $C_{12}$ and $C_{13}$. Said in other words, the non-affine correction decreases the affine elastic constants by about 70 $\%$ for $C_{11}$, $C_{33}$,$C_{44}$, $C_{66}$ and up to 90 $\%$ for $C_{12}$ and $C_{13}$.

\begin{table}[htbp]
\centering
\begin{tabular}{lcccccc}
\hline
Elast. Const. (GPa) &$C_{11}$ &$C_{33}$ &$C_{44}$ &$C_{66}$ &$C_{12}$ &$C_{13}$ \\ \hline
Affine+Nonaffine  &90.5 &107.0 &50.2 &41.1 &8.1 &15.2\\
Affine only &375.6 &329.6 &189.2 &125.4 &125.2 &189.1\\
Exp.~\cite{Will88} &86.8 &105.8 &58.2 &39.9 &7.0 &19.1\\
Exp.~\cite{Heyliger03} &87.3 &105.8 &57.2 & 40.4&6.57 &12.0\\
Exp.~\cite{Wang2015} &86.6 &106.4 &58 & &6.74 &12.4
\label{tab:elastconst}
\end{tabular}

\caption{Comparison between experimental measurements of the elastic constants of $\alpha$-quartz and the present numerical calculations, including both affine and nonaffine contributions or only the affine part.}

\end{table}

The nonaffine relaxations originate from the lack of symmetry of the $\alpha$-quartz crystal \cite{Milkus,Damart}. This is evident for the short-ranged pair potential part of the interatomic potential since in Eq. \ref{Xi_shortrange}, the nonaffine force vector, $\mathbf{\Xi}_i$, which drives the nonaffine relaxations, is written as a sum over  neighbors of terms of the type $D_{ij}^{\alpha \beta} R_{ij}^\kappa$ that add up to zero in a centro-symmetric environment. The same is true for the long-range terms in Eqs. \ref{tensileXi} and \ref{Xi_longrange_shear}, which depend on $\sum_j q_j \sin(\mathbf{G}.\mathbf{R}_{ij})$, which is also zero if atom $i$ is a center of centro-symmetry. In $\alpha$-quartz, neither Si nor O atoms are centers of symmetry, which may explain why nonaffine relaxations are so important in this crystal. However, Si atoms are surrounded by close-to-perfect tetrahedra of O atoms as explained in Sec. \ref{LattStruc}, while O atoms are in clearly asymmetrical environments since the Si-O-Si bonds are not straight, but make an angle close to 148$^o$. The higher symmetry of the environment of the Si atoms implies more limited nonaffine relaxations for these atoms. The latter depend on the imposed deformation, but we have checked that the nonaffine displacements of the Si atoms is systematically at least a factor of 2 smaller than that of the O atoms.

It was suggested in a recent work~\cite{Milkus} that the lack of centrosymmetry is responsible not only for the nonaffinity of the elastic constants, but also for the boson peak that shows up in the VDOS of glasses and non-centrosymmetric crystals. In ~\cite{Milkus}, model systems were studied numerically, which included random spring networks derived from glasses, and crystals with random bond-depletion. A universal correlation was found between the boson peak amplitude and a new order parameter for centrosymmetry (but importantly, \textit{not} with the standard bond-orientational order parameter), which allowed for the collapse of data from systems with very different lattice topologies (i.e. random networks and defective crystals).

The present findings demonstrate, for the first time, that strong nonaffine elasticity originates from non-centrosymmetry of the lattice also in \textit{perfectly-ordered} (defect-free) non-centrosymmetric crystals such as $\alpha$-quartz. Also in this case, the strong nonaffinity of the elastic constants is accompanied by a pronounced boson peak in the normalized VDOS, which shows up in both experimental measurements and atomic-scale simulations, in perfect mutual agreement as shown above in Fig. \ref{DOS}.

These observations rise the fundamental question about the microscopic mechanism which links the atomic-scale non-centrosymmetry of the lattice and the boson peak in the VDOS. In all systems studied so far, including the defective crystals of ~\cite{Milkus}, the boson peak frequency is very close to the frequency of the Ioffe-Regel crossover at which the phonons wavelength becomes smaller than their mean-free path and the phonons become quasi-localized. In glasses, this phenomenon is obviously driven by disorder, which is responsible for the scattering of the phonons on sufficiently small wavelengths. In a system like $\alpha$-quartz, it remains to be established whether non-centrosymmetry alone can induce similar scattering processes, which would lead to the peak. To elucidate this point, it will be necessary, in future work, to study more in detail the microscopics of the phonon propagation and how this is affected by non-centrosymmetry. For example, the non-centrosymmetry of the lattice has been recently shown to generate new physics in the phonon propagation, including chiral phonons~\cite{chiral_phonons}.



\section{Conclusion}

We have shown in the paper that employing an empirical potential for $\alpha$-quartz, with long-range Coulombic effects explicitly considered, the elastic constants of $\alpha$-quartz, consisting of the contribution of affine and nonaffine contributions, can be excellently recovered. It was found that the nonaffine force field in non-centro-symmetric lattice indeed plays a crucial role in the elastic constants. Here, we considered the static elastic constant, but the present framework can be readily extended to consider lattice dynamics at finite frequencies \cite{Damart}. Further, the dynamical structure factor can be easily calculated and qualitative comparison with experimental data will be a subject of a future study.

\begin{acknowledgements}
B.C. acknowledges the financial support from CSC-Cambridge Scholarship. D.R. acknowledges support from LABEX iMUST (ANR-10-LABX-0064) of Universit\'e de Lyon (programme Investissements d'Avenir, ANR-11-IDEX-0007).
\end{acknowledgements}

\bibliographystyle{unsrt}

\bibliography{biblio}

\end{document}